%% file: paper.tex
\documentclass[10pt,letterpaper,twocolumn,twoside]{IEEEtran}

\usepackage{amsfonts}
\usepackage{amsbsy}
\usepackage{amssymb}
\usepackage{amscd}
\usepackage[cmex10]{amsmath}
\usepackage{graphicx}
\usepackage[usenames,dvipsnames]{pstricks}
\usepackage{subfigure}
\usepackage[ruled,linesnumbered,vlined]{algorithm2e}
\usepackage[amsmath,thmmarks]{ntheorem}
\usepackage{wasysym}

\input{code}

\begin{document}

\title{Alternating Rate Profile Optimization \\ in Single Stream MIMO Interference Channels}

\author{Rami~Mochaourab,~%\IEEEmembership{Member,~IEEE,}
        Pan~Cao,~%\IEEEmembership{Student Member,~IEEE,}
        and~Eduard~Jorswieck,%~\IEEEmembership{Senior Member,~IEEE,}%
\thanks{\copyright ~2013 IEEE. Personal use of this material is permitted. Permission from IEEE must be obtained for all other uses, in any current or future media, including reprinting/republishing this material for advertising or promotional purposes, creating new collective works, for resale or redistribution to servers or lists, or reuse of any copyrighted component of this work in other works.}
\thanks{Part of this work was presented at the IEEE ICASSP, Vancouver, Canada, May 26–-31, 2013 \cite{Mochaourab2013}. Rami Mochaourab is with Fraunhofer Heinrich Hertz Institute, Berlin, Germany. Phone: +493031002392; Fax: +493031002250 E-mail: rami.mochaourab@ieee.org. Pan Cao and Eduard Jorswieck are with Communications Theory, Communications Lab, TU Dresden, Germany. (e-mail: pan.cao@tu-dresden.de, eduard.jorswieck@tu-dresden.de)}}%

\maketitle

\input{sections/abstract}

%\vspace{0.1cm}
%\noindent\textbf{EDICS: COM-MIMO; COM-CODPHY}
%\vspace{-0.25cm}
%
%
%\IEEEpeerreviewmaketitle
%
\input{sections/introduction}
\input{sections/system_model}
\input{sections/profile_opt}
\input{sections/alternating_opt}
\input{sections/conclusions}
\input{sections/appendix}

\bibliographystyle{IEEEbib}
\bibliography{references}

\end{document}

%% file: code.tex
\newtheorem{theorem}{Theorem}

\newtheorem{proposition}{Proposition}

\newcommand{\mat}[1]{\boldsymbol{#1}}

\newcommand{\pp}[1]{{\left( #1 \right)}}

\newcommand{\br}[1]{{\left\{ #1 \right\}}}
\newcommand{\norm}[1]{{ \left\Vert #1 \right\Vert }}
\newcommand{\abs}[1]{{ | #1 | }}
\newcommand{\sabs}[1]{{ | #1 |^2 }}
\newcommand{\snorm}[1]{{ \Vert #1 \Vert^2 }}

\DeclareMathOperator*{\maximize}{\text{maximize}}
\DeclareMathOperator*{\find}{\text{find}}

\def\miso{{\text{\tiny{MISO}}}}
\def\simo{{\text{\tiny{SIMO}}}}

\def\H{{{\dag}}} % Hermitian \dag
\def\bH{{\mat{H}}} % channel vector hh^H
\def\bI{{\mat{I}}} % Identity
\def\bQ{{\mat{Q}}} % transmit cov
\def\bz{{\mat{z}}} % vector z
\def\bw{{\mat{w}}} % beamforming vector w
\def\bv{{\mat{v}}} % beamforming vector v
\def\bg1{{g_{11}^{\perp}}}
\def\bg2{{g_{22}^{\perp}}}

\newcommand{\collec}[1]{\br{\mat{#1}}_{\setK}}%_{i\in \setK}

\def\setW{{\mathcal{W}}} % stable set
\def\setV{{\mathcal{V}}} % stable set
\def\setK{{\mathcal{K}}} % stable set

\def\rpa{{\text{rate-profile } \mat{\alpha}}}
\def\rpb{{\text{rate-profile } \mat{\alpha}'}}
\def\rpc{{\text{rate-profile}}}

%% file: sections/abstract.tex
\begin{abstract}
% setting
The multiple-input multiple-output interference channel is considered with perfect channel information at the transmitters and single-user decoding receivers. With all transmissions restricted to single stream beamforming, we consider the problem of finding all Pareto optimal rate-tuples in the achievable rate region. The problem is cast as a rate profile optimization problem. Due to its nonconvexity, we resort to an alternating approach: For fixed receivers, optimal transmission is known. For fixed transmitters, we show that optimal receive beamforming is a solution to an inverse field of values problem. We prove the solution's stationarity and compare it with existing approaches.%
\end{abstract}%
%This setting corresponds to the single-stream multiple-input multiple-output (MIMO) interference channel.
%An approach to attain different Pareto optimal points in the MIMO interference channel is rate profile optimization. %approach based on successive optimization of the transmit and receive beamforming vectors.

%% file: sections/introduction.tex
\section{Introduction}
% motivation on the problem
% optimality in multiuser systems and the problem and its complexity - The importance of the problem:
%
In multiuser networks, an efficient operating point is reached if it is not further possible to strictly improve the performance of all users jointly. Such an operating point is called Pareto optimal. The problem of characterizing all Pareto optimal operating points in the multiple-input multiple-output (MIMO) interference channel is still unsolved. It is known that finding specific Pareto optimal rate-tuples such as the maximum sum-rate, the proportional fair, and the max-min operating points are NP-hard problems \cite{Liu2013a}. The significance of solving these problems is to provide performance bounds for evaluating distributed low complexity algorithms, e.g. in \cite{Scutari2009a,Schmidt2013}.

% Accordingly, at a Pareto optimal operating point, the used communication resources are utilized efficiently.
% approaches to solve the problem - related work
% necessary condition and weighted sum rate - max-min
%
An approach to simplify the problem is to characterize the set of necessary transmission strategies for Pareto optimal operation for each user independently \cite{Park2012}. However, the search space remains very large for finding a Pareto optimal transmission strategy. Direct computation of Pareto optimal points can be performed through a maximization of the weighted sum-rate \cite{Negro2010a,Shi2011}. Alternating optimization algorithms to reach a local optimum are proposed for the MIMO interference channel in \cite{Negro2010a} and for the MIMO interfering broadcast channel in \cite{Shi2011}. The drawback of the weighted sum-rate approach is that it does not obtain points on the nonconvex part of the rate region \cite{Das1997}. To deal with this problem, max-min optimization is suitable and is considered using alternating optimization in \cite{Liu2013a} for the single-stream MIMO interference channel and in \cite{Razaviyayn2012} for the MIMO interfering broadcast channel. Both algorithms reach a local optimum of the original max-min problem. While max-min optimization can achieve points on the nonconvex parts of the rate region, it still fails to achieve all Pareto optimal points. In \cite{Cao2011}, a boundary intersection approach in the single stream MIMO interference channel is conducted using alternating optimization. A semi-definite relaxation is utilized to solve the problem of maximizing the rate of one user ensuring a fixed rate for the other users. %The convergent point is not guaranteed to be a stationary point. %In each alternation, the rate of one user is optimized ensuring a fixed rate for the other user.% The approach, however, is not extensible to more than two users.

% contributions: difference to max-min?
%
We consider the rate profile optimization problem for computing all Pareto optimal points in the single-stream MIMO interference channel rate region. Due to its nonconvexity, we adopt the common approach of alternating optimization. For fixed receiver beamforming, rate profile optimization is solved in \cite{Qiu2011}. For fixed transmit beamforming, the receive beamforming vectors are obtained by solving a set of feasibility problems each corresponding to an inverse field of value problem. With this respect, we reveal a link between the beamforming problem and a problem from matrix analysis. The advantage of our approach in comparison to max-min optimization is the explicit achievement of points along a deterministic rate profile ray. We prove the convergence of our solution to a stationary point of the original problem.

\emph{Notations:} Column vectors and matrices are given in lowercase and uppercase boldface letters, respectively. $\norm{\mat{a}}$ is the Euclidean norm of $\mat{a}\in \mathbb{C}^N$. $\abs{b}$ is the absolute value of $b\in\mathbb{C}$. $(\cdot)^\H$ denotes Hermitian transpose. $\mat{I}$ is an identity matrix. Define the collection $\collec{a} := (a_1,\ldots,a_\abs{\setK})$. $\mathcal{CN}(0,\mat{A})$ denotes a circularly-symmetric Gaussian complex random vector with covariance matrix $\mat{A}$.%

%% file: sections/system_model.tex
\section{System Model and Problem Formulation}\label{sec:sys_model}
Consider a set $\setK = \br{1,\ldots,K}$ of interfering links. Each transmitter $j$ uses $n_j$ antennas and each receiver $k$ uses $m_k$ antennas. The flat fading channel matrix from transmitter $j$ to receiver $k$ is $\bH_{jk} \in \mathbb{C}^{m_k \times n_j}$. We assume that each transmitter sends a single data stream to its intended receiver. The beamforming vector at transmitter $j$ is $\bw_j$ from the set
\begin{equation}\label{eq:tx_set}
\bw_j \in \setW_j = \br{\bw \in \mathbb{C}^{n_j}\mid\snorm{\bw} \leq 1},
\end{equation}
\noindent where we assumed a total transmit power constraint of one w.l.o.g. The received signal at receiver $k$ is $\mat{y}_k = \sum\nolimits_{j=1}^K \bH_{jk} \bw_j x_j + \bz_k,$ where $x_j\sim \mathcal{CN}(0,1)$ is the signal from transmitter $j$ and $\bz_k \sim \mathcal{CN}(0,\bI \sigma^2)$ is additive white Gaussian noise. Assuming single-user decoding, the achievable rate of link $k$ after equalization with $\bv_k$ is
\begin{equation}
%\label{eq:Rate}
R_k(\bv_k,\collec{w}) = \text{C}\pp{\frac{\sabs{\bv_k^\H \bH_{kk} \bw_k}}{\sigma^2 \snorm{\bv_k} + \sum\nolimits_{j\neq k} \sabs{\bv_k^\H \bH_{jk} \bw_j}}},\hspace{-0.05cm}
\end{equation}
\noindent with $\text{C}(x) = \log_2(1 + x)$. Since the user rate is not affected by receiver power, the beamforming vector, $\bv_k$ is chosen from
\begin{equation}\label{eq:rx_set}
\bv_k \in \setV_k = \br{\bv \in \mathbb{C}^{m_k}\mid\snorm{\bv} = 1},
\end{equation}
\noindent where the receive power is normalized to one. The rate region
%\begin{equation}\label{eq:rate_region}
%\mathcal{R}=\{\mat{r} \in \mathbb{R}_+^K\mid r_k = R_k(\bv_k,\collec{w}), \bw_k \in \setW_k, \bv_k\in \setV_k, k \in \setK\},
%\end{equation}
\begin{multline}\label{eq:rate_region}
\mathcal{R}=\{\mat{r} \in \mathbb{R}_+^K\mid r_k = R_k(\bv_k,\collec{w}), \bw_k \in \setW_k,\\ \bv_k\in \setV_k, k \in \setK\},
\end{multline}
\noindent is the $K$-dimensional set composed of all rate tuples. The set of Pareto optimal points in $\mathcal{R}$ is defined as \cite[p. 14]{Peters1992}:
\begin{equation}\label{eq:wPareto_boundary}
\mathcal{W}(\mathcal{R})=\br{\mat{x} \in \mathcal{R}|\text{\small{there is no }} \mat{y} \in \mathcal{R} \text{\small{ with }} \mat{y} > \mat{x}},
\end{equation}
\noindent with componentwise inequality in \eqref{eq:wPareto_boundary}. At a Pareto optimal point it is impossible to strictly improve the performance of all users jointly. The set of \emph{strong} Pareto optimal points is a subset of $\mathcal{W}(\mathcal{R})$ defined as:
\begin{equation}\label{eq:Pareto_boundary}
\mathcal{P}(\mathcal{R})=\br{\mat{x} \in \mathcal{R}|\text{\small{there is no }} \mat{y} \in \mathcal{R} \text{\small{ with }} \mat{y} \geq \mat{x}, \mat{y} \neq \mat{x}},\hspace{-0.15cm}
\end{equation}
\noindent where the inequality in \eqref{eq:Pareto_boundary} is componentwise.

In this work, we are interested in characterizing all points in $\mathcal{W}(\mathcal{R})$ in \eqref{eq:wPareto_boundary}, the so-called \emph{Pareto boundary} of the rate region $\mathcal{R}$. Any Pareto optimal point is attained as a solution of the following \emph{rate profile optimization problem}:
\begin{subequations}\label{eq:original_prob}
\begin{eqnarray} \label{eq:original_prob1}
\maximize_{\collec{v},\collec{w},{R}} & & {R}\\ \label{eq:original_prob2}
s.t. & & R_k(\bv_k,\collec{w}) \geq \alpha_k {R}, ~~ k \in \setK,\\ \label{eq:original_prob3}
& & \bw_k \in \setW_k, ~~ \bv_k \in \setV_k, ~~ k \in \setK.
\end{eqnarray}
\end{subequations}
Note that the rate profile approach has been first proposed for broadcast and multiple-access channels in \cite{Mohseni2006} and for MISO interference channels in \cite{Zhang2010}. In \eqref{eq:original_prob}, the rate profile $(\alpha_1,\ldots,\alpha_K)$ satisfies $\alpha_k \geq 0, k \in \setK$ and $\sum_{k=1}^K \alpha_k = 1$. The objective ${R}$ corresponds to the links' sum-rate if the constraints in \eqref{eq:original_prob3} are satisfied with equality. The rate profile defines the direction of a ray starting in the origin of the rate region, and the point of intersection of the ray and the Pareto boundary is a solution of \eqref{eq:original_prob}. Solving \eqref{eq:original_prob} for all possible rate profiles, all points in $\mathcal{W}(\mathcal{R})$ are characterized. Problem \eqref{eq:original_prob} is however nonconvex and even NP-hard \cite{Liu2013a}, and hence no method is known that can attain its solution efficiently.

We propose to decompose problem \eqref{eq:original_prob} into two subproblems which are solved alternatingly. The first problem optimizes the transmit beamforming vectors for fixed receive beamforming vectors. The second problem optimizes the receive beamforming vectors for fixed transmit beamforming vectors. Next, we discuss the two problems independently. Later in Section \ref{sec:alternating}, the solutions of the two problems are used to construct the alternating algorithm.%

%% file: sections/profile_opt.tex
\section{Optimality in MISO and SIMO Channels}\label{sec:optimality}
\subsection{Rate Profile Optimization in MISO Interference Channels}\label{sec:MISO}
In this section, we assume the receive beamforming vectors are fixed. The considered MIMO setting reduces to a MISO interference channel, and the rate region is a subset of $\mathcal{R}$:
\begin{equation}
{\mathcal{R}}^\miso=\{\mat{r} \in \mathcal{R}|r_k=R_k(\bv_k,\collec{w}), \bw_k \in \setW_k, k \in \setK\}.%\hspace{-0.18cm}
\end{equation}
\noindent For fixed receive beamforming $\collec{v}$, problem \eqref{eq:original_prob} reduces to%
\begin{subequations}\label{eq:problem_miso}
\begin{eqnarray}
\maximize_{\collec{w},{R}} & & {R}\\ \label{eq:constraint_MISO}
s.t. & & R_k(\bv_k,\collec{w}) \geq \alpha_k {R}, ~~ k \in \setK,\\
& & \bw_k \in \setW_k, ~~ k \in \setK.
\end{eqnarray}
\end{subequations}
\noindent It is shown in \cite{Qiu2011} that problem \eqref{eq:problem_miso} can be solved by a set of feasibility problems:
\begin{subequations}\label{eq:problem_miso_feas}
\begin{eqnarray}
\find & & \bw_1,\ldots,\bw_K \\
s.t. & & R_k(\bv_k,\collec{w}) \geq \alpha_k t, ~~ k \in \setK,\\
& & \bw_k \in \setW_k, ~~ k \in \setK,
\end{eqnarray}
\end{subequations}
\noindent where the parameter $t>0$ is updated based on a bisection method. In order to determine the feasibility, the problem in \eqref{eq:problem_miso_feas} is transformed in \cite[Section II.D]{Qiu2011} to a second order cone programm (SOCP) and solved efficiently.

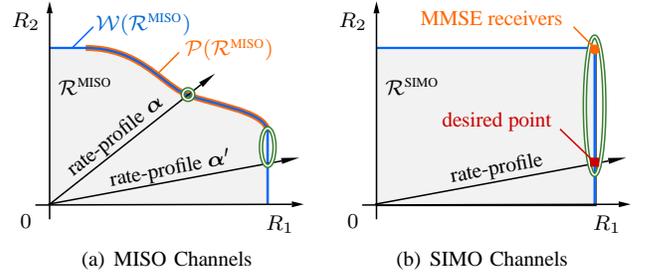
\begin{figure}[t]
\centering
%\begin{pspicture}(0.5,-1.5)(4.2,1.5)
\subfigure[\label{fig:illust_opt1} MISO Channels]{{\input{figures/RP_MISO2}}}
%\begin{pspicture}(0.3,-1.5)(4,1.5)
\subfigure[\label{fig:illust_opt2} SIMO Channels]{{\input{figures/RP_SIMO2}}}
\caption{\label{fig:illust_opt}Illustration of the solutions of rate profile optimization.}
\end{figure}

In \figurename~\ref{fig:illust_opt1}, solutions of problem \eqref{eq:problem_miso} are illustrated. For a rate profile ray $\mat{\alpha}$ passing through the set $\mathcal{P}(\mathcal{R}^\miso)$ (according to \eqref{eq:Pareto_boundary}), the solution of \eqref{eq:problem_miso} achieves a unique point in the rate region. If the rate profile ray does not pass through $\mathcal{P}(\mathcal{R}^\miso)$ as rate profile ray $\mat{\alpha}'$, then multiple solutions for \eqref{eq:problem_miso} exist corresponding to the points in the illustrated larger circle.
\subsection{Rate Profile Optimization in SIMO Interference Channels}\label{sec:SIMO}
In this section, we assume the transmit beamforming vectors $\collec{w}$ are fixed. The setting corresponds to a SIMO interference channel. The rate region in the SIMO setting is a subset of the rate region $\mathcal{R}$ in \eqref{eq:rate_region} and has the following property.
\begin{proposition}\label{thm:SIMO}
The rate region of a SIMO interference channel with fixed transmitters is a $K$-dimensional box:
\begin{equation}
\mathcal{R}^\simo = \{ \mat{r} \in \mathcal{R}: r_k \leq R_k({\bv}^*_k,\collec{w}), k \in \setK\}, \text{ with}
\end{equation}
\noindent
\begin{equation}\label{eq:mmse_beam}
{\bv}^*_k = \frac{\big(\sigma^2 \bI + \sum\nolimits_{j\neq k} \bH_{jk} \bw_j \bw_j^\H \bH_{jk}^\H \big)^{-1} \bH_{kk} \bw_k}{\big\| {\big({\sigma^2 \bI + \sum\nolimits_{j\neq k} \bH_{jk} \bw_j \bw_j^\H \bH_{jk}^\H \big) }^{-1} \bH_{kk} \bw_k}\big\|}.
\end{equation}
\end{proposition}
\begin{IEEEproof}
The proof can be found in \cite[Appendix A]{Mochaourab2013}.
\end{IEEEproof}
In \figurename~\ref{fig:illust_opt2}, an illustration of a two-user SIMO rate region is given. A single strong Pareto optimal point exists corresponding to joint minimum mean square error (MMSE) receive beamforming in \eqref{eq:mmse_beam}. The rate profile optimization for fixed transmit beamforming vectors is formulated as:
\begin{subequations}\label{eq:SIMO_opt1}
\begin{eqnarray}
\maximize_{\collec{v},{R}} & & {R}\\ \label{eq:SIMO_const}
s.t. & & R_k(\bv_k,\collec{w}) \geq \alpha_k {R}, ~~ k \in \setK,\\
& & \bv_k \in \setV_k, ~~ k \in \setK.
\end{eqnarray}
\end{subequations}
\noindent The receive beamforming vector that optimize \eqref{eq:SIMO_opt1} is not necessarily unique. In \figurename~\ref{fig:illust_opt2}, an illustration of the set of points that solve \eqref{eq:SIMO_opt1} are contained in the green circle. One solution of \eqref{eq:SIMO_opt1} is joint MMSE beamforming. Another special solution corresponds to the intersection of the rate profile ray and the Pareto boundary. Note that rate profile optimization in the SIMO interference channel has been considered in \cite[Section IV.B]{Liu2012} and \cite{Liu2013}. In comparison, we do not optimize the transmission power but only the receive beamforming vectors by a different approach.

In order to attain the desired point on the rate profile ray, we can solve the following set of feasibility problems:
\begin{subequations}\label{eq:feasibility_simo}
\begin{eqnarray}
\find & & \bv_1,\ldots,\bv_K\\ \label{eq:SIMO_constraint}
s.t. & & R_k(\bv_k,\collec{w}) = \alpha_k t, ~~ k \in \setK,\\
& & \bv_k \in \setV_k, ~~ k \in \setK.
\end{eqnarray}
\end{subequations}
\noindent where $t \geq 0$ is updated according to a bisection method. In comparison to \eqref{eq:SIMO_opt1}, the inequality in \eqref{eq:SIMO_const} is changed to equality in \eqref{eq:SIMO_constraint}. We can reformulate \eqref{eq:SIMO_constraint} to
\begin{equation}\label{eq:constraint_SIMO2}
\bv_k^\H \bQ_k(t) \bv_k = 0, ~~ k \in \setK,% \text{ with}
\end{equation}
\noindent where the Hermitian matrix $\bQ_k(t) = \bH_{kk} \bw_k \bw_k^\H \bH_{kk}^\H - \pp{2^{\alpha_k t} - 1} \big({\sigma^2 \bI + \sum\nolimits_{j\neq k} \bH_{jk} \bw_j \bw_j^\H \bH_{jk}^\H}\big).$ The problem in \eqref{eq:feasibility_simo} with \eqref{eq:SIMO_const} replaced by \eqref{eq:constraint_SIMO2} is called the \emph{inverse field of values problem} \cite{Carden2009}. In order to check the feasibility of \eqref{eq:feasibility_simo} for a chosen $t$, it suffices to test whether $0$ lies between the smallest and largest eigenvalues of $\bQ_k(t)$, i.e., $0$ is in the field of values \cite[Chapter 1]{Horn1991} of $\bQ_k(t)$. After the convergence of the bisection method which determines the optimal $t$, each vector $\bv_k$ is determined by the algorithm from \cite{Carden2009} requiring five eigenvalue decompositions \cite[Section 5]{Carden2009}.%
%\cite{Carden2011}.%
%:$\mathcal{F}(\bX) = \br{ \bx^\H \bX \bx \in \mathbb{R} \mid \snorm{\bx} = 1 }$. Testing whether $0 \in \mathcal{F}(\bQ_k(t))$ is equivalent to checking whether zero lies between the smallest and largest eigenvalues of $\bQ_k(t)$. The field of values $\mathcal{F}(\bX)$ is a compact convex set. If $\bX$ is Hermitian, then $\mathcal{F}(\bX) \subset \mathbb{R}$ with the smallest element and largest element corresponding to the smallest and largest eigenvalues of the matrix $\bX$, respectively.
%

%% file: figures/RP_MISO2.tex
% Generated with LaTeXDraw 2.0.8
% Sun Jun 23 10:03:35 CEST 2013
% \usepackage[usenames,dvipsnames]{pstricks}
% \usepackage{epsfig}
% \usepackage{pst-grad} % For gradients
% \usepackage{pst-plot} % For axes
\scalebox{1} % Change this value to rescale the drawing.
{
\begin{pspicture}(0.5,-1.5)(4.2,1.5)
\definecolor{color24b}{rgb}{0.9490196078431372,0.9490196078431372,0.9490196078431372}
\definecolor{color24}{rgb}{0.8,0.8,0.8}
\definecolor{color1}{rgb}{0.0,0.40784313725490196,0.8}
\definecolor{color2}{rgb}{1.0,0.40784313725490196,0.0}
\definecolor{color3058}{rgb}{0.0,0.4,1.0}
\definecolor{color3083}{rgb}{0.0,0.40784313725490196,1.0}
\pspolygon[linewidth=0.002,linecolor=color24,fillstyle=solid,fillcolor=color24b](0.8476562,0.98)(0.8476562,-1.1)(3.7276561,-1.1)(3.7276561,-0.1)(1.3476562,0.98)
\usefont{T1}{ptm}{m}{n}
\rput(2.0967188,1.295){\footnotesize \color{color1}
 $\mathcal{W}(\mathcal{R}^\miso)$}
\usefont{T1}{ptm}{m}{n}
\rput(3.2367187,0.955){\footnotesize \color{color2}
 $\mathcal{P}(\mathcal{R}^\miso)$}
\usefont{T1}{ptm}{m}{n}
\rput(1.32125,0.475){\footnotesize $\mathcal{R}^\miso$}
\psbezier[linewidth=0.08,linecolor=color2,fillstyle=solid,fillcolor=color24b](1.3276561,0.98)(2.0133228,0.98)(2.3153229,0.5400383)(2.6076562,0.38945454)(2.8999896,0.23887081)(3.747656,0.15473685)(3.747656,-0.14)
\psbezier[linewidth=0.03,linecolor=color3058](1.3276561,0.98)(2.0133228,0.98)(2.3153229,0.5400383)(2.6076562,0.38945454)(2.8999896,0.23887081)(3.747656,0.15473685)(3.747656,-0.14)
\psline[linewidth=0.02cm,arrowsize=0.04cm 2.0,arrowlength=3.0,arrowinset=0.0]{->}(0.8476562,-1.1)(3.0476563,0.64)
\pscircle[linewidth=0.02,linecolor=OliveGreen,dimen=outer,doubleline=true,doublesep=0.02,doublecolor=white](2.6876562,0.36){0.1}
%\psdots[dotsize=0.08](0.8476562,-1.1)
\usefont{T1}{ptm}{m}{n}
\rput(0.53124994,-1.325){\footnotesize $0$}
\psline[linewidth=0.03cm,linecolor=color3058](0.8476562,0.98)(1.3476562,0.98)
\psline[linewidth=0.03cm,linecolor=color3058](3.747656,-1.1)(3.747656,-0.12)
\usefont{T1}{ptm}{m}{n}
\rput(0.53124994,1.335){\footnotesize $R_2$}
\usefont{T1}{ptm}{m}{n}
\rput(3.9112499,-1.365){\footnotesize $R_1$}
%\psframe[linewidth=0.02,dimen=outer](4.447656,1.8)(0.0476562,-1.8)
\psline[linewidth=0.02cm,arrowsize=0.04cm 2.0,arrowlength=3.0,arrowinset=0.0]{->}(0.8276562,-1.1)(4.1676564,-0.48)
\psline[linewidth=0.02cm,arrowsize=0.04cm 2.0,arrowlength=2.0,arrowinset=0.0]{->}(0.8476562,-1.3)(0.8476562,1.6)
\psline[linewidth=0.02cm,arrowsize=0.04cm 2.0,arrowlength=2.0,arrowinset=0.0]{->}(0.6576562,-1.11)(4.2276564,-1.1)
\usefont{T1}{ptm}{m}{n}
\rput{10.495942}(-0.075007856,-0.46212795){\rput(2.46125,-0.625){\footnotesize $\rpb$}}
\usefont{T1}{ptm}{m}{n}
\rput{38.567}(0.25700536,-1.1333202){\rput(1.7312499,-0.185){\footnotesize $\rpa$}}
\psellipse[linewidth=0.02,linecolor=OliveGreen,dimen=outer,doubleline=true,doublesep=0.02,doublecolor=white](3.747656,-0.34)(0.12,0.28)
\psline[linewidth=0.02cm,linecolor=color3083](1.1462349,0.9785786)(1.4290775,1.2614213)
\psline[linewidth=0.02cm,linecolor=color2](2.2862349,0.63857865)(2.5690775,0.92142135)
\end{pspicture}
}

%% file: figures/RP_SIMO2.tex
% Generated with LaTeXDraw 2.0.8
% Sun Jun 23 10:31:03 CEST 2013
% \usepackage[usenames,dvipsnames]{pstricks}
% \usepackage{epsfig}
% \usepackage{pst-grad} % For gradients
% \usepackage{pst-plot} % For axes
\scalebox{1} % Change this value to rescale the drawing.
{
\begin{pspicture}(0.3,-1.5)(4,1.5)
\definecolor{color773}{rgb}{0.8,0.0,0.0}
\definecolor{color482b}{rgb}{0.9490196078431372,0.9490196078431372,0.9490196078431372}
\definecolor{color488}{rgb}{0.0,0.4,1.0}
\definecolor{color528}{rgb}{1.0,0.4,0.0}
\definecolor{color501}{rgb}{1.0,0.40784313725490196,0.0}
\psframe[linewidth=0.002,dimen=outer,fillstyle=solid,fillcolor=color482b](3.723125,0.98)(0.803125,-1.12)
\usefont{T1}{ptm}{m}{n}
\rput(1.2767187,0.475){\footnotesize $\mathcal{R}^\simo$}
%\psdots[dotsize=0.08](0.803125,-1.1)
\usefont{T1}{ptm}{m}{n}
\rput(0.48671874,-1.325){\footnotesize $0$}
\psline[linewidth=0.03cm,linecolor=color488](0.803125,0.98)(3.703125,0.98)
\psline[linewidth=0.03cm,linecolor=color488](3.703125,-1.1)(3.703125,0.98)
\usefont{T1}{ptm}{m}{n}
\rput(0.48671874,1.335){\footnotesize $R_2$}
\usefont{T1}{ptm}{m}{n}
\rput(3.8667188,-1.365){\footnotesize $R_1$}
%\psframe[linewidth=0.02,dimen=outer](4.403125,1.8)(0.003125,-1.8)
\psline[linewidth=0.02cm,arrowsize=0.04cm 2.0,arrowlength=3.0,arrowinset=0.0]{->}(0.783125,-1.1)(4.123125,-0.48)
\psline[linewidth=0.02cm,arrowsize=0.04cm 2.0,arrowlength=2.0,arrowinset=0.0]{->}(0.803125,-1.3)(0.803125,1.6)
\psline[linewidth=0.02cm,arrowsize=0.04cm 2.0,arrowlength=2.0,arrowinset=0.0]{->}(0.613125,-1.11)(4.183125,-1.1)
\usefont{T1}{ptm}{m}{n}
\rput{10.495942}(-0.075920284,-0.45219424){\rput(2.4067187,-0.625){\footnotesize $\rpc$}}
\usefont{T1}{ptm}{m}{n}
\rput(2.3371875,1.375){\footnotesize \color{color528}MMSE receivers}
\psellipse[linewidth=0.02,linecolor=OliveGreen,dimen=outer,doubleline=true,doublesep=0.02,doublecolor=white](3.703125,0.21)(0.16,0.95)
\pscircle[linewidth=0.002,linecolor=color528,dimen=outer,fillstyle=solid,fillcolor=color528](3.703125,0.96){0.06}
\psline[linewidth=0.02cm,linecolor=color501](3.4217036,1.2614213)(3.7045465,0.9785786)
\usefont{T1}{ptm}{m}{n}
\rput(2.4025,-0.005){\footnotesize \color{color773}desired point}
\psline[linewidth=0.02cm,linecolor=color773](3.243125,-0.1)(3.7245464,-0.56142133)
\psframe[linewidth=0.02,linecolor=color773,dimen=outer,fillstyle=solid,fillcolor=color773](3.763125,-0.48)(3.643125,-0.6)
\end{pspicture}
}

%% file: sections/alternating_opt.tex
\section{Algorithm and Numerical Results}\label{sec:alternating}
\begin{algorithm}[t]
\KwIn{rate profile $\collec{\alpha} = (\alpha_1,\ldots,\alpha_K)$ and accuracy $\epsilon$}
\textbf{Initialize}: $i=0;$ choose random $\collec{v}^{{{(0)}}}$\;
\Repeat{${R}^{(i)} - {R}^{(i-1)} < \epsilon$}{
solve \eqref{eq:problem_miso} given $\collec{v}^{{{(i)}}}$ to get $\big(\collec{w}^{{{(i+1)}}},{R}^{(i+1)}\big)$\;
solve \eqref{eq:SIMO_opt1} given $\collec{w}^{{{(i+1)}}}$ to get $\big(\collec{v}^{{{(i+2)}}},{R}^{{{(i+2)}}}\big)$\;
$i = i + 2$\;
}
\KwOut{$\collec{w}^{{{(i)}}},\collec{v}^{{{(i)}}}$ }
\caption{\label{alg:alt_prof}Alternating rate profile optimization.}
\end{algorithm}
The alternating rate profile algorithm is described in Algorithm \ref{alg:alt_prof}. The measure ${R}^{(i)}$ at iteration $i$ is the achieved progress from the origin along the rate profile ray. In each iteration $i$, an improvement ${R}^{(i)} - {R}^{(i-1)} \geq \epsilon > 0$ must be achieved.% The algorithm terminates if the improvement is less than $\epsilon$.% Since the rate region is a bounded set, the alternating algorithm is guaranteed to converge according to the monotone convergence theorem.
\begin{theorem}\label{thm:stationarity}
The alternating rate profile optimization in Algorithm \ref{alg:alt_prof} converges to a stationary point of \eqref{eq:original_prob}.
\end{theorem}
\begin{IEEEproof}
The proof is provided in Appendix \ref{proof:stationarity}.
\end{IEEEproof}

\begin{figure}[t]
  \centering
  \includegraphics[width=\linewidth,clip]{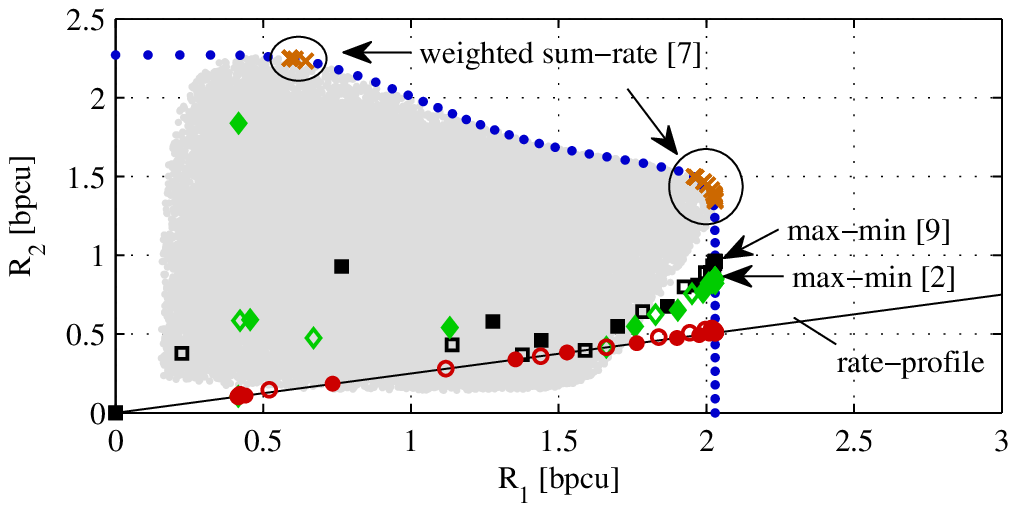}
  %\vspace{-0.4cm}
  \caption{Plot of a two-user rate region at signal-to-noise ratio SNR$ = 1/\sigma^2 =0$ dB and two antennas at each transmitter and receiver. The points marked with $\square$, $\lozenge$, and $\Circle$ correspond to the rate tuples achieved in each iteration of the alternating algorithms in \cite{Razaviyayn2012}, \cite{Liu2013a}, and Algorithm \ref{alg:alt_prof}, respectively. The filled (unfilled) markers correspond to receiver (transmitter) optimization. }\label{fig:RR2D} %\vspace{-0.2cm}
\end{figure}

In \figurename~\ref{fig:RR2D}, a two-user rate region is plotted. Single random channel matrix realisations are selected. The cloud of points in \figurename~\ref{fig:RR2D} corresponds to random norm-one transmit beamforming vectors with MMSE receive beamforming. Using Algorithm~\ref{alg:alt_prof} we are able to plot for $50$ different rate profile samples the points bounding the rate region. In \figurename~\ref{fig:RR2D}, the points marked with cross correspond to the algorithm in \cite{Shi2011} where an iterative weighted MMSE algorithm is proposed to optimize the weighted sum-rate. It can be observed that points on the nonconvex part of the Pareto boundary are not achieved.

\begin{figure}[t]
  \centering
  \includegraphics[width=\linewidth,clip]{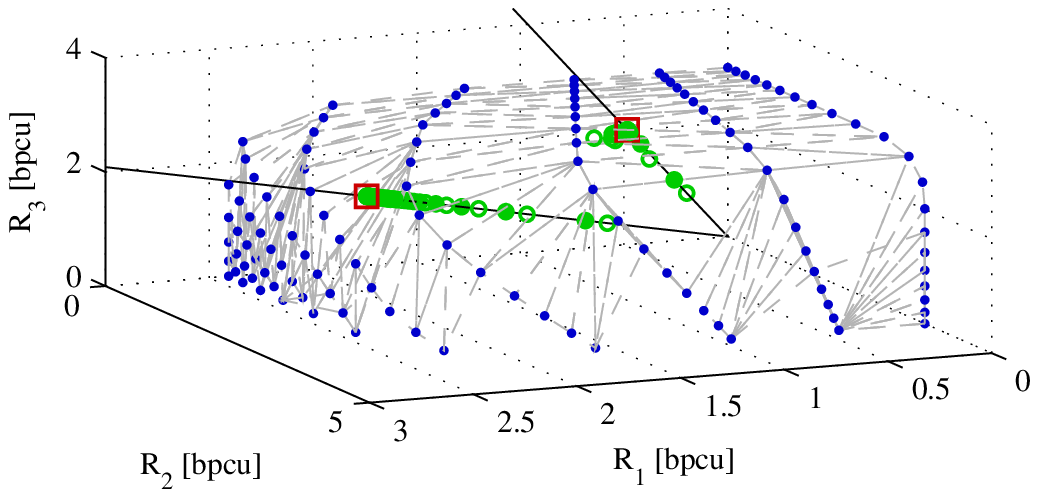}
  %\vspace{-0.4cm}
  \caption{Plot of three-user rate region at SNR$=0$ dB and three antennas at each transmitter and receiver. For the two rate profiles, the filled (unfilled) markers correspond to receiver (transmitter) optimization of Algorithm~\ref{alg:alt_prof}.}\label{fig:RR3D} %\vspace{-0.2cm}
  \vspace{-0.5cm}
\end{figure}
For a selected rate profile not passing through a strong Pareto optimal point, the solutions of the transmitter and receiver optimizations are plotted during the alternating optimization. The performance improvement in each iteration can be observed and the alternating optimization terminates at a point very close to the Pareto boundary. In comparison to max-min optimization in \cite{Liu2013a} and \cite{Razaviyayn2012}, Algorithm~\ref{alg:alt_prof} delivers a solution on the rate profile ray. The difference between Algorithm 1 and \cite[Algorithm ECCAA]{Liu2013a} is the receiver optimization step where in \cite{Liu2013a} MMSE receive beamforming is performed. While the algorithm in \cite{Razaviyayn2012} does not terminate at the rate profile ray, it is powerful enough to solve the max-min problem for the general setting with multiple data streams per user and multiple users associated with each transmitter. Note that both algorithms in \cite{Liu2013a} and \cite{Razaviyayn2012} terminate at the rate profile ray if it passes through a strong Pareto optimal point. Generally, it is hard to anticipate whether the rate profile ray passes through a strong Pareto optimal point. This can be observed in a three-user rate region in \figurename~\ref{fig:RR3D}. The terminating points of Algorithm~\ref{alg:alt_prof}, marked with a square, always achieve points on the rate profile ray.

%% file: sections/conclusions.tex
\section{Conclusions}\label{sec:conc}
We have considered rate profile optimization in the single stream MIMO interference channel in order to characterize all Pareto optimal points in the rate region. Due to the nonconvexity of the problem, we have choosen an alternating optimization approach. For fixed receivers, we use an existing method for rate profile optimization in MISO channels. For fixed transmitters, we have shown that rate profile optimization can be solved by a set of feasibility problems each corresponding to an inverse field of value problem. In comparison to existing algorithms, we always achieve points along the rate profile ray. We prove that the proposed solution is a stationary point of the original problem.

%% file: sections/appendix.tex
\appendices
\section{Proof of Theorem \ref{thm:stationarity}}\label{proof:stationarity}
Denote the optimization of Problem \eqref{eq:problem_miso} and the optimization of Problem \eqref{eq:SIMO_opt1} by the function $\{\mat{w}\}_{\mathcal{K}} = \Psi(\{\mat{v}\}_{\mathcal{K}})$ and $\{\mat{v}\}_{\mathcal{K}} = \Theta(\{\mat{w}\}_{\mathcal{K}})$, respectively.
In Algorithm~\ref{alg:alt_prof}, the sequence $\big\{{R}(\{\mat{w}\}_{\mathcal{K}}^{(i)}, \{\mat{v}\}_{\mathcal{K}}^{(i)})\big\}_{i=1}^{\infty}$ \emph{monotonically} increases as the iteration number $i$ increases due to the optimality of $\Theta(\cdot)$ and $\Psi(\cdot)$, and additionally is upper-bounded. The convergence of $\big\{{R}(\{\mat{w}\}_{\mathcal{K}}^{(i)}, \{\mat{v}\}_{\mathcal{K}}^{(i)})\big\}_{i=1}^{\infty}$ and thus the convergence of Algorithm~\ref{alg:alt_prof} is guaranteed. %for any feasible initial point $\{\mat{w}\}_{\mathcal{K}}^{(0)}$.

Let $\lim_{i \rightarrow \infty}{R}(\{\mat{w}\}_{\mathcal{K}}^{(i)}, \{\mat{v}\}_{\mathcal{K}}^{(i)})
\stackrel{\Delta}{=} \widehat{R}(\{\widehat{\mat{w}}\}_{\mathcal{K}}, \{\widehat{\mat{v}}\}_{\mathcal{K}})$ denote the convergent point\footnote{There must exist a cluster point, denoted by $\{\widehat{\mat{w}}\}_{\mathcal{K}}$, of $\big\{\{\mat{w}\}_{\mathcal{K}}^{(i)}\big\}_{i=1}^{\infty}$ due to the compactness of the set of $\{\mat{w}\}_{\mathcal{K}}$, and the limit of $\big\{\{\mat{v}\}_{\mathcal{K}}^{(i)}\big\}_{i=1}^{\infty}$ can be expressed as $\Theta\left(\{\widehat{\mat{w}}\}_{\mathcal{K}}\right)$ because $\Theta(\cdot)$ is a continuous function.}. It remains to show that $\left(\{\widehat{\mat{w}}\}_{\mathcal{K}}, \{\widehat{\mat{v}}\}_{\mathcal{K}}\right)=\left(\{\widehat{\mat{w}}\}_{\mathcal{K}}, \Theta\left(\{\widehat{\mat{v}}\}_{\mathcal{K}}\right)\right)$ is a stationary solution to Problem \eqref{eq:original_prob}. Assume that $\left({R}^*, \{{\mat{w}}^*\}_{\mathcal{K}}, \{{\mat{v}}^*\}_{\mathcal{K}}\right)$ associated with Lagrange multipliers $(\{\mu_k^*\}_{\mathcal{K}},\{\zeta_k^*\}_{\mathcal{K}}, \{\eta_k^*\}_{\mathcal{K}})$ is a stationary solution to Problem \eqref{eq:original_prob}, which must satisfy the following KKT conditions of Problem \eqref{eq:original_prob}:
\begin{subequations}
\begin{align}
%&\nabla_{{R}}\mathcal{L}({R}, \{\mat{w}\}_{\mathcal{K}}, \{\mat{v}\}_{\mathcal{K}}) =
1 - \sum\limits_{k \in \mathcal{K}}\mu_k^*\alpha_k&=0, \label{eq:KKT11} \\
%&\nabla_{\mat{w}_k}\mathcal{L}({R}, \{\mat{w}\}_{\mathcal{K}}, \{\mat{v}\}_{\mathcal{K}}) =
\sum\limits_{k \in \mathcal{K}}\mu_k^* \nabla_{\mat{w}_k}R_k(\mat{v}_k^*, \{\mat{w}^*\}_{\mathcal{K}}) - 2\zeta_k^*\mat{w}_k^* &=0,~~ \forall k\in \mathcal{K}, \label{eq:KKT12} \\
%&\nabla_{\mat{v}_k}\mathcal{L}({R}, \{\mat{w}\}_{\mathcal{K}}, \{\mat{v}\}_{\mathcal{K}}) =
\mu_k^* \nabla_{\mat{v}_k}R_k(\mat{v}_k^*, \{\mat{w}^*\}_{\mathcal{K}}) - 2\eta_k^*\mat{v}_k^* &=0,~~ \forall k\in \mathcal{K}, \label{eq:KKT13}\\
0\leq \mu_k^* \perp R_k(\mat{v}_k^*, \{\mat{w}^*\}_{\mathcal{K}}) - \alpha_k {R}^* &\geq 0,~~ \forall k\in \mathcal{K}, \label{eq:KKT14} \\
0\leq \zeta_k^* \perp 1-\mat{w}_k^{*,\H}\mat{w}_k^* & \geq 0,~~ \forall k\in \mathcal{K}, \label{eq:KKT15} \\
0< \eta_k^*,~~\mat{v}_k^{*,\H}\mat{v}_k^* &=1,~~ \forall k\in \mathcal{K}. \label{eq:KKT16}
\end{align}\end{subequations}

Given $\{{\mat{v}}\}_{\mathcal{K}} = \{\widehat{\mat{v}}\}_{\mathcal{K}}$, it is clear $\big(\widehat{R}, \{\widehat{\mat{w}}\}_{\mathcal{K}}=\Theta(\{\widehat{\mat{v}}\}_{\mathcal{K}})\big)$ is the optimal solution to Problem \eqref{eq:problem_miso}. Therefore, $\big(\widehat{R}, \{\widehat{\mat{w}}\}_{\mathcal{K}}\big)$ associated with Lagrange multipliers $(\{\widehat{\mu}_k\}_{\mathcal{K}},\{\widehat{\zeta}_k\}_{\mathcal{K}})$ must satisfy the following KKT conditions of Problem \eqref{eq:problem_miso}:
\begin{subequations}
\begin{align}
1 - \sum\limits_{k \in \mathcal{K}}\widehat{\mu}_k\alpha_k &=0, \label{eq:KKT31} \\
\sum\limits_{k \in \mathcal{K}}\widehat{\mu}_k \nabla_{\mat{w}_k}R_k(\widehat{\mat{v}}_k, \{\widehat{\mat{w}}\}_{\mathcal{K}}) - 2\widehat{\zeta}_k\widehat{\mat{w}}_k&=0,~~ \forall k\in \mathcal{K}, \label{eq:KKT32} \\
0\leq \widehat{\mu}_k \perp R_k(\widehat{\mat{v}}_k, \{\widehat{\mat{w}}\}_{\mathcal{K}}) - \alpha_k \widehat{R} & \geq 0,~~ \forall k\in \mathcal{K}, \label{eq:KKT33} \\
0\leq \widehat{\zeta}_k \perp 1-\widehat{\mat{w}}_k^{\H}\widehat{\mat{w}}_k &\geq 0,~~ \forall k\in \mathcal{K}. \label{eq:KKT34}
\end{align}\end{subequations}

Similarly, $\{\widehat{\mat{v}}\}_{\mathcal{K}}=\Psi(\{\widehat{\mat{w}}\}_{\mathcal{K}})$ in Problem \eqref{eq:SIMO_opt1} corresponds the following KKT conditions \eqref{eq:KKT31}, \eqref{eq:KKT33} and
\begin{subequations}
\begin{align}
%&1 - \sum_{\mathcal{K}}\widetilde{\mu}_k\alpha_k =0; \label{eq:KKT41} \\
\widehat{\mu}_k \nabla_{\mat{v}_k}R_k(\widehat{\mat{v}}_k, \{\widehat{\mat{w}}\}_{\mathcal{K}}) -2\widehat{\eta}_k \widehat{\mat{v}}_k&=0,~~ \forall k\in \mathcal{K}, \label{eq:KKT41} \\
0< \widehat{\eta}_k,~~\widehat{\mat{v}}_k^{\H}\widehat{\mat{v}}_k&=1,~~ \forall k\in \mathcal{K}. \label{eq:KKT42}
\end{align}\end{subequations}

Combining the KKT conditions \eqref{eq:KKT31}-\eqref{eq:KKT34} of Problem \eqref{eq:problem_miso} and \eqref{eq:KKT41}-\eqref{eq:KKT42} of Problem \eqref{eq:SIMO_opt1} and comparing with the KKT conditions \eqref{eq:KKT11}-\eqref{eq:KKT16}, we have that $\big(\widehat{R}, \{\widehat{\mat{w}}\}_{\mathcal{K}}, \{\widehat{\mat{v}}\}_{\mathcal{K}}\big)$ associated with the Lagrange multipliers $(\{\widehat{\mu}_k, \widehat{\zeta}_k, \widehat{\eta}_k\}_{\mathcal{K}}$ satisfy the KKT conditions of Problem \eqref{eq:original_prob}, i.e., \eqref{eq:KKT11}-\eqref{eq:KKT16}. It implies that $\big(\widehat{R}, \{\widehat{\mat{w}}\}_{\mathcal{K}}, \{\widehat{\mat{v}}\}_{\mathcal{K}}\big)$ is a stationary solution to Problem \eqref{eq:original_prob}.%
%(9)\eqref{eq:problem_miso}
%(13)\eqref{eq:SIMO_opt1}
%(7)\eqref{eq:original_prob}